\documentclass[final,5p,times,preprint]{elsarticle}

\usepackage{amsfonts}
\usepackage{amsmath}
\usepackage{amssymb}
\usepackage{bm}
\usepackage{comment}
\usepackage{xcolor}
\usepackage{booktabs}
\usepackage[normalem]{ulem}
\usepackage[export]{adjustbox}
\graphicspath{{./figures/}}
\journal{Physics Letters B}

\usepackage[utf8]{inputenc}
\usepackage[english]{babel}
\biboptions{comma,sort&compress}

\usepackage{amsmath}
\usepackage{graphicx}
\usepackage[colorinlistoftodos]{todonotes}
\usepackage[colorlinks=true, allcolors=blue]{hyperref}
\usepackage[normalem]{ulem} 

\def\nuc#1#2{\relax\ifmmode{}^{#1}{\protect\text{#2}}\else${}^{#1}$#2\fi}

\begin{document}

\begin{frontmatter}

\title{Structure of $^{13}$Be probed via quasi-free scattering}

\author[cea]{A.~Corsi}
\cortext[mail]{Corresponding author}
\ead{acorsi@cea.fr}
\author[rik,cns,tuda]{Y.~Kubota}
\author[ECT,pd]{J. Casal}
\author[FAMN]{M.~Gómez-Ramos}
\author[FAMN]{A.~M.~ Moro}
\author[cea3]{G. Authelet}
\author[rik]{H. Baba}
\author[tuda]{C. Caesar}
\author[cea2]{D. Calvet}
\author[cea2]{A. Delbart}
\author[cns]{M. Dozono}
\author[pek]{J. Feng}
\author[ipno]{F. Flavigny}
\author[cea3]{J.-M. Gheller}
\author[lpc]{J. Gibelin}
\author[cea2]{A. Giganon}
\author[cea]{A. Gillibert}
\author[toh]{K. Hasegawa}
\author[rik]{T. Isobe}
\author[miy]{Y. Kanaya}
\author[miy]{S. Kawakami}
\author[ehw]{D. Kim}
\author[cns]{Y. Kiyokawa}
\author[cns]{M. Kobayashi}
\author[tod]{N. Kobayashi}
\author[toh]{T. Kobayashi}
\author[tit]{Y. Kondo}
\author[rik]{Z. Korkulu}
\author[tod]{S. Koyama}
\author[cea]{V. Lapoux}
\author[miy]{Y. Maeda}
\author[lpc]{F. M. Marqu\'es}
\author[rik]{T. Motobayashi}
\author[tod]{T. Miyazaki}
\author[tit]{T. Nakamura}
\author[kyo]{N. Nakatsuka}
\author[kyu]{Y. Nishio}
\author[cea,tuda]{A. Obertelli}
\author[kyu]{A. Ohkura}
\author[lpc]{N. A. Orr}
\author[cns]{S. Ota}
\author[rik]{H. Otsu}
\author[tit]{T. Ozaki}
\author[rik,cea]{V. Panin}
\author[tuda]{S. Paschalis}
\author[cea]{E. C. Pollacco}
\author[tum]{S. Reichert}
\author[cea4]{J.-Y. Rousse}
\author[tit]{A. T. Saito}
\author[kyu]{S. Sakaguchi}
\author[rik]{M. Sako}
\author[cea]{C. Santamaria}
\author[rik]{M. Sasano}
\author[rik]{H. Sato}
\author[tit]{M. Shikata}
\author[rik]{Y. Shimizu}
\author[kyu]{Y. Shindo}
\author[rik]{L. Stuhl}
\author[rik]{T. Sumikama}
\author[cea,tuda]{Y.L. Sun}
\author[kyu]{M. Tabata}
\author[tit]{Y. Togano}
\author[tit]{J. Tsubota}
\author[rik]{T. Uesaka}
\author[rik]{Z. H. Yang}
\author[kyu]{J. Yasuda}
\author[rik]{K. Yoneda}
\author[rik]{J. Zenihiro}

\address[cea]{D\'epartement de Physique Nucl\'eaire, IRFU, CEA, Universit\'e Paris-Saclay, F-91191 Gif-sur-Yvette, France}
\address[rik]{RIKEN Nishina Center, Hirosawa 2-1, Wako, Saitama 351-0198, Japan}

\address[cea2]{D\'epartement d'\'electronique des D\'etecteurs et d'Informatique pour la Physique, IRFU, CEA, Universit\'e Paris-Saclay, F-91191 Gif-sur-Yvette, France}
\address[cea3]{	D\'epartement des Acc\'el\'erateurs, de Cryog\'enie et de Magn\'etisme, IRFU, CEA, Universit\'e Paris-Saclay, F-91191 Gif-sur-Yvette, France}
\address[cea4]{D\'epartement d'Ing\'enierie des Syst\`emes, IRFU, CEA, Universit\'e Paris-Saclay, F-91191 Gif-sur-Yvette, France}

\address[cns]{Center for Nuclear Study, University of Tokyo, Hongo 7-3-1, Bunkyo, Tokyo 113-0033, Japan}

\address[ECT]{European Centre for Theoretical Studies in Nuclear Physics and Related Areas (ECT$^*$), Villa Tambosi, Strada delle Tabarelle 286, I-38123 Trento, Italy}
\address[pd]{Dipartimento di Fisica e Astronomia ``G.~Galilei'' and INFN - Sezione di Padova, Via Marzolo 8, I-35131 Padova, Italy}
\address[FAMN]{Departamento de F\'{\i}sica At\'omica, Molecular y Nuclear, Facultad de F\'{\i}sica, Universidad de Sevilla, Apartado 1065, E-41080 Sevilla, Spain}

\address[tuda]{Department of Physics, Institut für Kernphysik, Technische Universität Darmstadt, 64289 Darmstadt, Germany}
\address[pek]{Department of Physics, Peking University, China}
\address[ipno]{Institut de Physique Nucleaire Orsay, IN2P3-CNRS, F-91406 Orsay Cedex, France}
\address[lpc]{LPC Caen, ENSICAEN, Universite de Caen, CNRS/IN2P3, F-14050 Caen, France}
\address[toh]{Department of Physics, Tohoku University, Aramaki Aza-Aoba 6-3, Aoba, Sendai, Miyagi 980-8578, Japan}
\address[miy]{Department of Applied Physics, University of Miyazaki, Gakuen-Kibanadai-Nishi 1-1, Miyazaki 889-2192, Japan}
\address[ehw]{Department of Physics, Ehwa Womans University}
\address[tod]{Department of Physics, University of Tokyo, Hongo 7-3-1, Bunkyo, Tokyo 113-0033, Japan}
\address[tit]{Department of Physics, Tokyo Institute of Technology, 2-12-1 O-Okayama, Meguro, Tokyo 152-8551, Japan}
\address[kyo]{Department of Physics, Kyoto University, Kitashirakawa, Sakyo, Kyoto 606-8502, Japan}
\address[kyu]{Department of Physics, Kyushu University, Nishi, Fukuoka 819-0395, Japan}
\address[tum]{Department of Physics, Technische Universitat Munchen, Germany}


\begin{abstract}
We present an investigation of the structure of $^{13}$Be obtained via a kinematically complete measurement of the $(p,pn)$ reaction in inverse kinematics at 265 MeV/nucleon. 
The relative energy spectrum of $^{13}$Be is compared to Transfer-to-the-Continuum calculations which use as structure inputs the overlaps of the  $^{14}$Be ground-state wave function, computed in a three-body model, with the unbound states of the $^{13}$Be residual nucleus. The key role of neutron $p$-wave orbital in the interpretation of the low-relative-energy part of the spectrum is discussed.
\end{abstract}


\end{frontmatter}

\section{Introduction \label{sec:intro} }
Light nuclei at the neutron dripline display exotic properties connected to the large spatial distribution of weakly bound valence neutrons, such as halo and clustering. A special class of halo nuclei are two-neutron halo nuclei like $^6$He, $^{11}$Li and $^{14}$Be, known as Borromean nuclei, since they can be described as a three-body system without any bound two-body subsystem. They  provide a good environment to study dineutron correlations which are expected to be a key element in their stabilization \cite{sag15}. Different methods have been advocated and used to probe the properties of Borromean nuclei, such as Coulomb dissociation \cite{nak06,aum13}, dineutron decay \cite{hag15}, and quasi-free scattering reactions \cite{kik16}. As mentioned, the subsystems constituted of one neutron and the remaining fragment ($^5$He, $^{10}$Li, $^{13}$Be) are unbound and their continuum exhibit a resonant structure. 

In this paper we present a study of the spectroscopy of the unbound $^{13}$Be nucleus obtained by measuring the invariant mass of the $^{12}$Be-neutron system resulting from the decay of the $^{13}$Be system produced via the quasi-free scattering reaction $^{14}$Be($p$,$pn$). 
The unbound nature of $^{13}$Be was suggested more than 30 years ago \cite{pos66, art70}, and confirmed in 1973 \cite{bow73}. Several experiments have attempted to study the spectroscopy of $^{13}$Be both via missing mass and invariant mass technique using charge exchange \cite{mar15}, fragmentation \cite{tho00}, proton removal from $^{14}$B \cite{lec04,ran14,rib18} and neutron removal from $^{14}$Be \cite{sim07,kon10,aks13a,aks13b}. The missing mass technique offers the advantage of yielding the absolute energy above the one-neutron decay threshold, and typically allows to explore a larger range in excitation energy above the two-neutron separation threshold. Using this method, resonances at $\sim$2, 5, 7, and 10~MeV were observed in Ref.~\cite{kor95}, and at 1.22(10), 2.10(16), 4.14(12), 5.09(14), and 7.0(2)~MeV in \cite{bel98}. Invariant mass spectra from different experiments display a peak at about 0.5~MeV above the $^{12}$Be+neutron threshold, and a broader structure around 2~MeV. As already discussed in Ref.~\cite{nak17}, the spectral shape strongly differs depending on the production mechanism, namely if $^{13}$Be is produced starting from $^{14}$Be \cite{sim07,kon10, aks13a,aks13b} or $^{14}$B \cite{lec04,ran14,rib18}. Even limiting ourselves to the first case, the mechanism adopted in this work, different interpretations of the relative energy spectrum have been provided. Ref.~\cite{kon10} interprets the low lying peak as a  1/2$^-$ ($\ell$=1) intruder state that appears due to the quenching of the $N$=8 spin-orbit shell gap, and the structure around 2~MeV as a  5/2$^+$ ($\ell$=2) state. This interpretation is based on the analysis of the transverse momentum distribution using $s$, $p$ and $d$ waves, corroborated by shell-model calculations, and is in agreement with predictions by \cite{bla07}. Ref.~\cite{aks13b} makes a synthesis of existing experimental results, with special emphasis on those obtained from proton-induced one-neutron removal \cite{kon10}. Nevertheless, the analysis of the transverse momentum distribution performed in Ref.~\cite{aks13b} yields quite different conclusions with respect to Ref.~\cite{kon10}: a much stronger $d$-wave component (dominant above 2.6~MeV), and a dominance of $s$-wave (80(10)\%) around 0.5~MeV, instead of $p$-wave. 

This diversity and sometimes inconsistency in the positions and spin assignment of the states of $^{13}$Be indicate that the standard fitting procedures used for the analysis of these spectra may be lacking some constraints on the possible structures due to the complexity of $^{13}$Be spectrum. As such, in this work we study the $^{14}$Be$(p,pn)$ reaction using a novel method, proposed in \citep{gomezramosplb17}, which uses consistent two- and three-body models for $^{14}$Be and $^{13}$Be and is able to provide predictions for the positions and weights of the structures of the spectrum, thus reducing the ambiguities in the analysis.

Part of the complexity of the $^{13}$Be continuum spectrum stems from the admixtures of single-particle structures with core-excited components. In fact, core excitation has been postulated as a key element to understand the formation of Borromean systems \cite{tan08,pot10}, but it is very difficult to pin down. The level scheme of $^{12}$Be is well established. A strong excitation of the $^{12}$Be(2$^+$) state in inelastic proton scattering, consistent with a strong quadrupole deformation, provided a first evidence of $N=8$ shell gap quenching in $^{12}$Be \cite{iwa00}. Furthermore, neutron removal from $^{12}$Be revealed that the last neutrons have a significant ($2s_{1/2}$)$^2$ +($1d_{5/2}$)$^2$ configuration and that there is only of order 30\% of the $(1p_{1/2})^2$ closed shell component \cite{pai06}.
In this experiment we were able to measure with high statistics the possible $^{12}$Be(2$^+$, 1$^-$) core excited component that decays via gamma rays. 

The experiment and the results on the spectroscopy of $^{13}$Be are presented in Sec.~\ref{sec:level2}, while their interpretation follows in Sec.~\ref{sec:level4} after a brief description of the theoretical framework.

\section{\label{sec:level2} Experiment}

\begin{figure}[t]
\begin{center}
      \fbox{\includegraphics[trim={16cm 8cm 3cm 5cm},clip,width=0.37\textwidth]{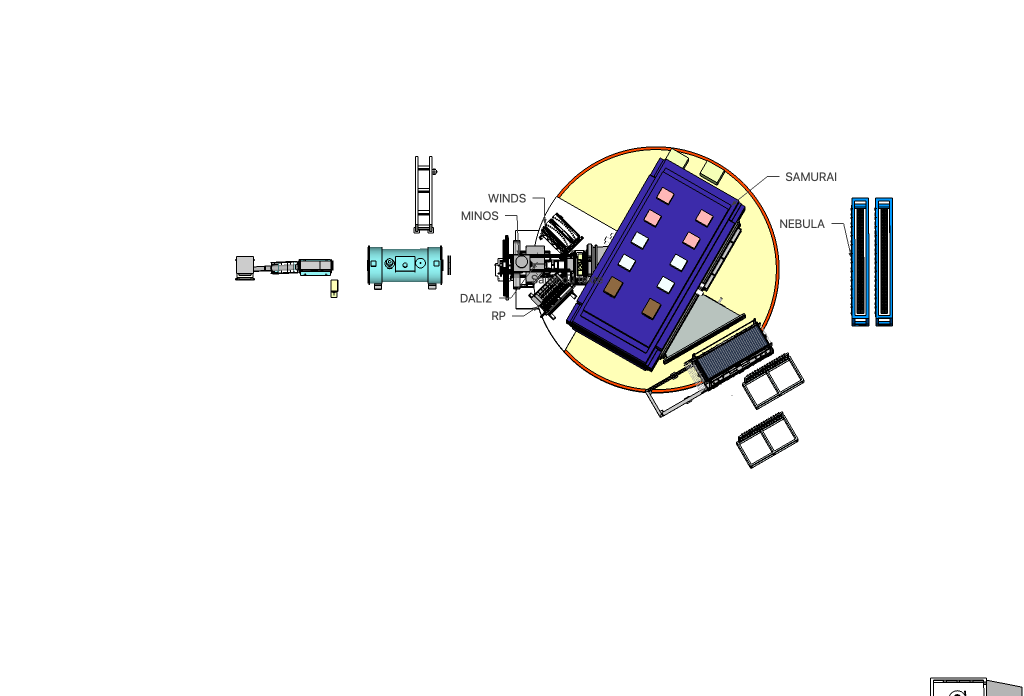}}
    \caption{\label{f:stp} Scheme of the experimental setup. } 
\end{center}
\end{figure}

\begin{figure}[t]
\begin{center}
      \includegraphics[trim={0cm -1cm 0.cm 0cm},clip,width=0.21\textwidth]{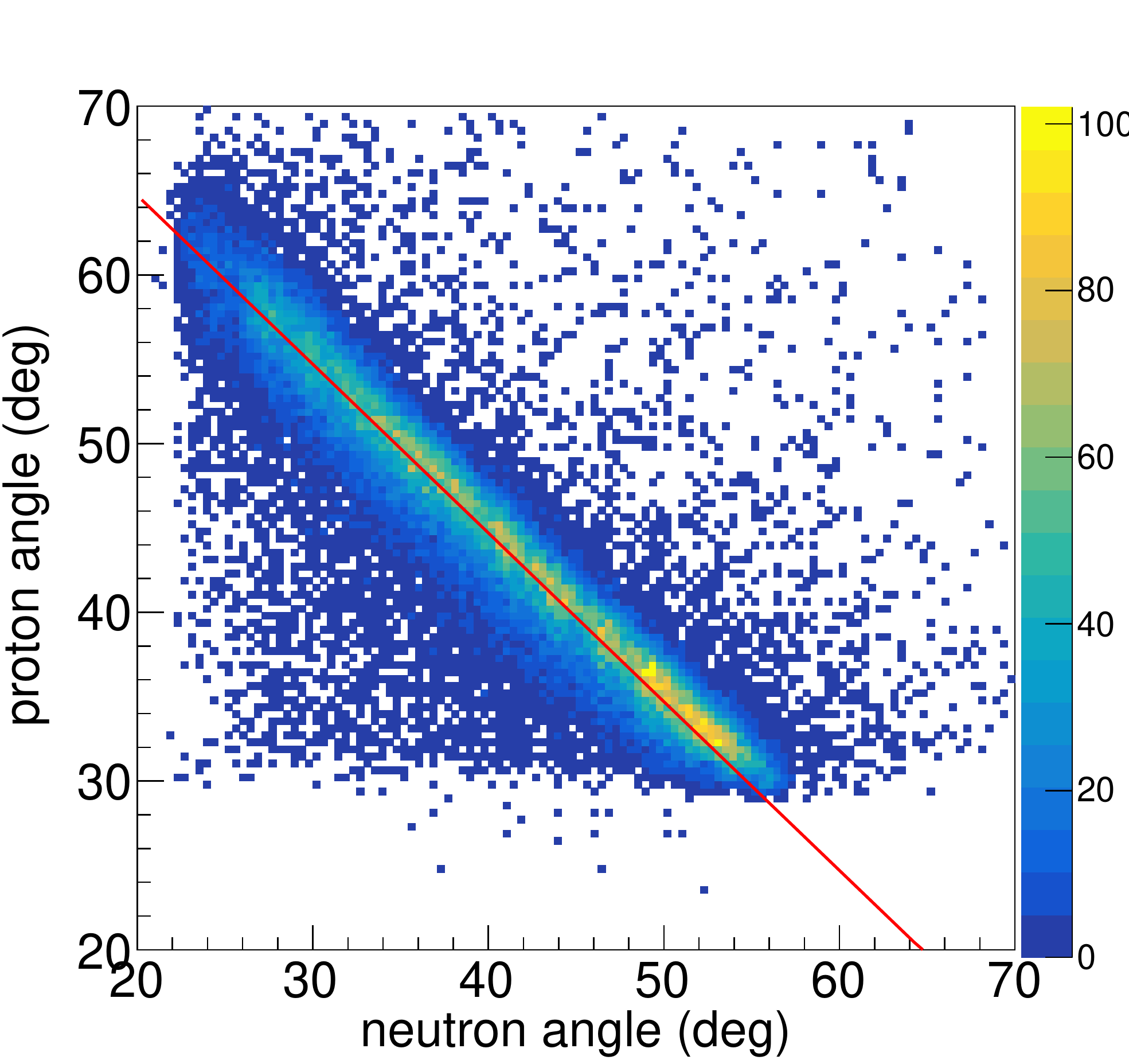}
      \includegraphics[trim={1.7cm 8.3cm 5cm 8cm},clip,width=0.21\textwidth]{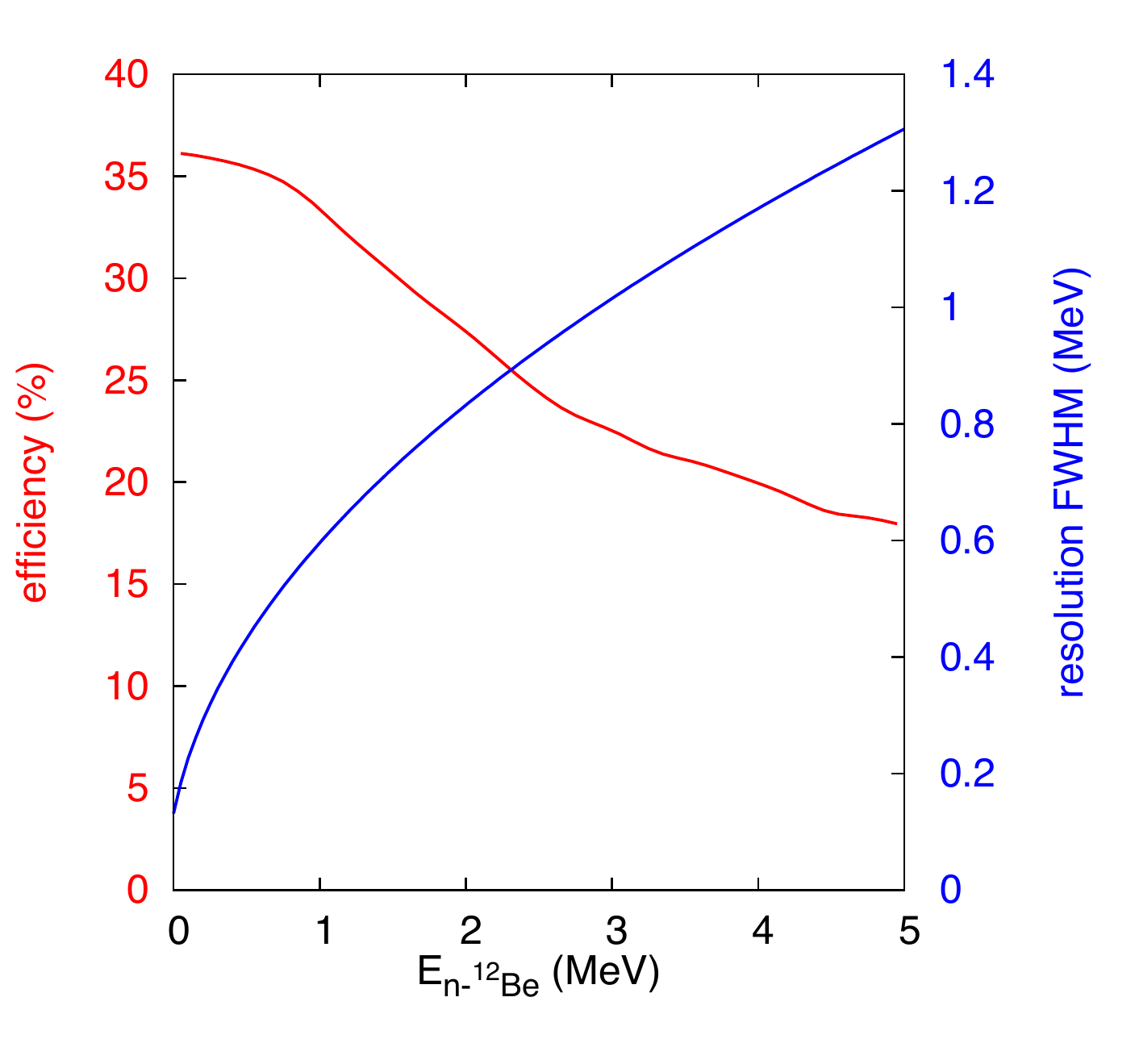}
    \caption{\label{f:qfs} Left: angular correlation in the laboratory frame of proton and neutron in $^{14}$Be$(p,pn)$ reaction. The red line correspond to the kinematics calculated with the QFS code of Ref. \cite{pan16}. Right: efficiency (red line) and resolution (blue line) for $^{13}$Be invariant mass measurement with NEBULA and SAMURAI. } 
\end{center}
\end{figure}

The experiment was performed at the Radioactive Isotope Beam Factory operated by the RIKEN Nishina Center and the Center for Nuclear Study (CNS) of the University of Tokyo. Secondary
beams were produced and separated by the BigRIPS fragment separator \cite{fuk13}, using projectile fragmentation of a $^{48}$Ca primary beam at 345 MeV/nucleon with a typical intensity of 400 particle nA on a Be target. Fragmentation products were detected and identified by using plastic scintillators and multi-wire drift chambers (MWDCs) positioned along the BigRIPS line.
The main components of the cocktail beams were $^{11}$Li, $^{14}$Be, and $^{17}$B (80\%, 12\%, and 8\%, respectively) and impinged on the secondary target with an average energy of 246, 265 and 277 MeV/nucleon, respectively. The secondary target was a 15-cm thick liquid hydrogen target surrounded by a Time Projection Chamber (TPC) of the MINOS device \cite{obe14}. The TPC in conjunction with the beam tracking MWDC detectors was acting as vertex tracker, allowing to improve invariant-mass and gamma spectroscopy resolution. The combination of the thick MINOS target and the intense secondary beams from RIBF ($\sim$ 10$^5$ pps) was a key ingredient to obtain enough statistics for a kinematically complete measurement.
A rather complex detection system (Fig. \ref{f:stp}) was deployed to perform an exclusive measurement.
The knocked-out neutron was measured by the WINDS array of plastic scintillators \cite{yak12}. Its kinetic energy was deduced with the time of flight technique. The recoil proton was tracked first in the TPC, then in a MWDC, and subsequently traversed an array of plastic scintillators allowing the measurement of its kinetic energy via the time of flight technique. The ensemble of MWDC and plastic scintillators wall is called the recoil proton (RP) detector hereafter. The identification and momentum analysis of the heavy charged fragment was achieved via the combination of tracking in the SAMURAI \cite{kob13} dipole magnet via a set of MWDC placed before and after the magnet, and the energy loss and time of flight measurement in an array of plastic scintillators placed at the focal plane of SAMURAI. Their momentum could then be deduced from the measurement of their trajectory. The dipole gap was kept under vacuum using a chamber equipped with thin exit windows \cite{shi13} so as to reduce to a minimum the amount of material encountered by both the fragments and neutrons. The decay neutron was detected in the two walls of plastic scintillators of the NEBULA array \cite{nak16}. The efficiency of the NEBULA array for the detection of one neutron is $\sim$ 35$\%$ at the beam energy of this experiment. The WINDS and recoil proton detector covered angles between 20$^\circ$ and 60$^\circ$, and 30$^\circ$ to 65$^\circ$, respectively, allowing the selection of high-momentum-transfer events corresponding to quasi-free scattering \cite{mar66,mar73,aum13}. 
In this process, the dominant mechanism for the knockout reaction is a single interaction between the incident particle and the struck nucleon which yields a kinematics for the $(p,pn)$ reaction very close to the one of free scattering. The angular correlation between the scattered neutron and the recoil proton is shown in Fig.~\ref{f:qfs} (left). The opening angle in the laboratory frame corresponds to 85$^\circ$, close to 86$^\circ$ as predicted using the QFS kinematics simulation of Ref. \cite{pan16}. This confirms that the high-momentum transfer ($>$~1 fm$^{-1}$) events selected by the detection system correspond to QFS. Invariant-mass resolution and efficiency have been determined via a GEANT4 simulation \cite{geant} with the code used in \cite{nak16}. The resolution follows a FWHM=$0.587\sqrt{E_r}$ MeV law and is shown in Fig.~\ref{f:qfs} (right). The efficiency is also shown in Fig.~\ref{f:qfs} (right) and is estimated assuming 100\% transmission of the beam and the fragment. The transmission (including tracking detectors efficiency and loss of beam and fragment in the thick MINOS target) is evaluated separately from experimental data taking the average of the values obtained for $^{12}$Be and $^{14}$Be beam transmission (65.0(1)\% and 62.9(2)\%, respectively), and corresponds to 64(1)\%. 


$^{12}$Be core excited states that decay via gamma emission (2$^+$, 1$^-$) were identified using a reduced version of the DALI2 gamma array consisting of 68 crystals, partially covering angles between 34$^\circ$ and 115$^\circ$ and arranged in order to avoid interference with $(p,pn)$ measurement. Photopeak efficiency of this reduced version of DALI2 (called DALI2 hereafter) was 8.9(5)\% and 7.0(4)\% at 2.1 and 2.7 MeV, respectively. This experiment was not sensitive to $^{12}$Be(0$^+_2$) core excited state as this state is isomeric and its gamma-decay lifetime (1.87(15) $\mu$s) is too long for in-flight detection \cite{shi07}. 

\begin{figure}[h!]
 \begin{center}
       \includegraphics[width=1\columnwidth]{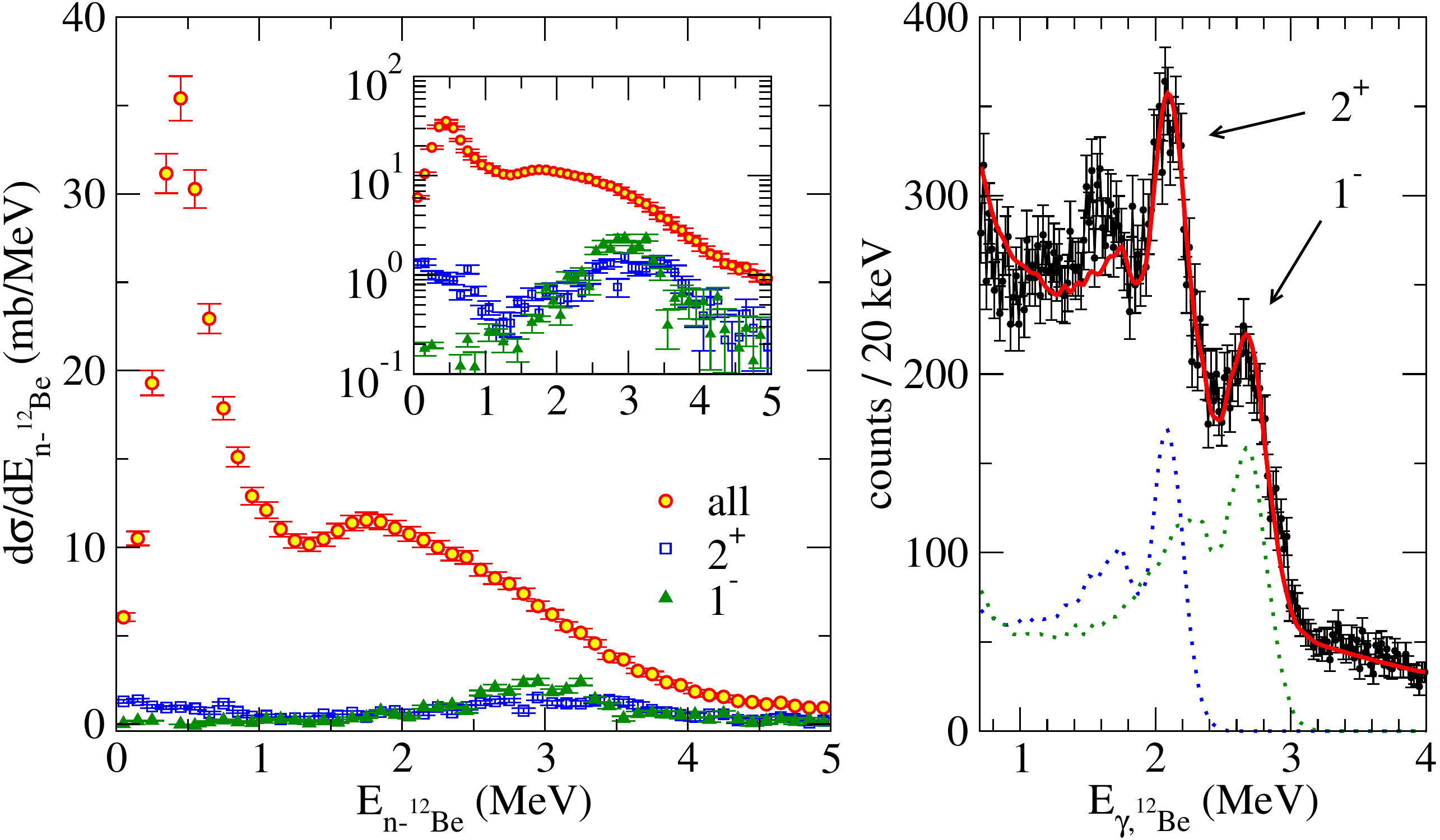}
     \caption{\label{f:erel} Left: relative energy spectrum of $^{13}$Be and contributions from core excited components. The inset shows the spectrum in logarithmic scale. Right: gamma spectrum of $^{12}$Be. The two transitions are reproduced by the sum of an exponential background and the response functions (dashed curves) of DALI2 to a transition at 2.1 MeV and 2.7 MeV, obtained via a GEANT4 simulation.} 
 \end{center}
 \end{figure}

 The invariant mass spectrum of $^{13}$Be is shown in Fig.~\ref{f:erel} (left). The absolute cross section is determined taking into account the efficiency for invariant mass measurement and the fragment transmission. The error bars take into account the uncertainty on the transmission (1\%), on the neutron detection efficiency (2.5\%) and the statistical uncertainty on the number of beam and fragment particles. The spectrum is characterized by a prominent peak with maximum at $\sim$~0.48~MeV and a broader structure, peaked at $\sim$~2.3 MeV, extending from $\sim$1~MeV to $\sim$5~MeV. 
 The contribution corresponding to $^{12}$Be(2$^+$) and $^{12}$Be(1$^-$) core excited states has been fixed via coincidences with 2.1 and 2.7 MeV gamma transitions, respectively, and is shown for comparison after correcting for gamma-detection efficiency. The uncertainty on gamma detection efficiency (6\%) has been added to the error bars.

 The gamma spectrum of $^{12}$Be is shown in Fig.~\ref{f:erel} (right). The 2.1(0.1) and 2.7(0.4) MeV transitions are consistent with the known transitions deexciting the 2$^+$ and 1$^-$ excited states of $^{12}$Be to its ground state. We note that the same gamma transitions were observed in Ref.~\cite{kon10}, though with very limited statistics, while  \cite{rib18} observed only the 2.1~MeV transition. As can be better seen in the inset of Fig. \ref{f:erel} (left), the 2.1~MeV one is observed in coincidence with a structure peaking at $\sim$0 and $\sim$3 MeV in the relative energy spectrum, as in \cite{rib18}. The 2.7~MeV one is observed in coincidence with a structure at $\sim$3 MeV. The contribution from the Compton events associated to the 2.7~MeV transition summing up to the 2.1~MeV transition has been estimated via the simulation and subtracted from the cross section.  

\begin{figure}[h!]
 \begin{center}
       \includegraphics[width=0.8\columnwidth]{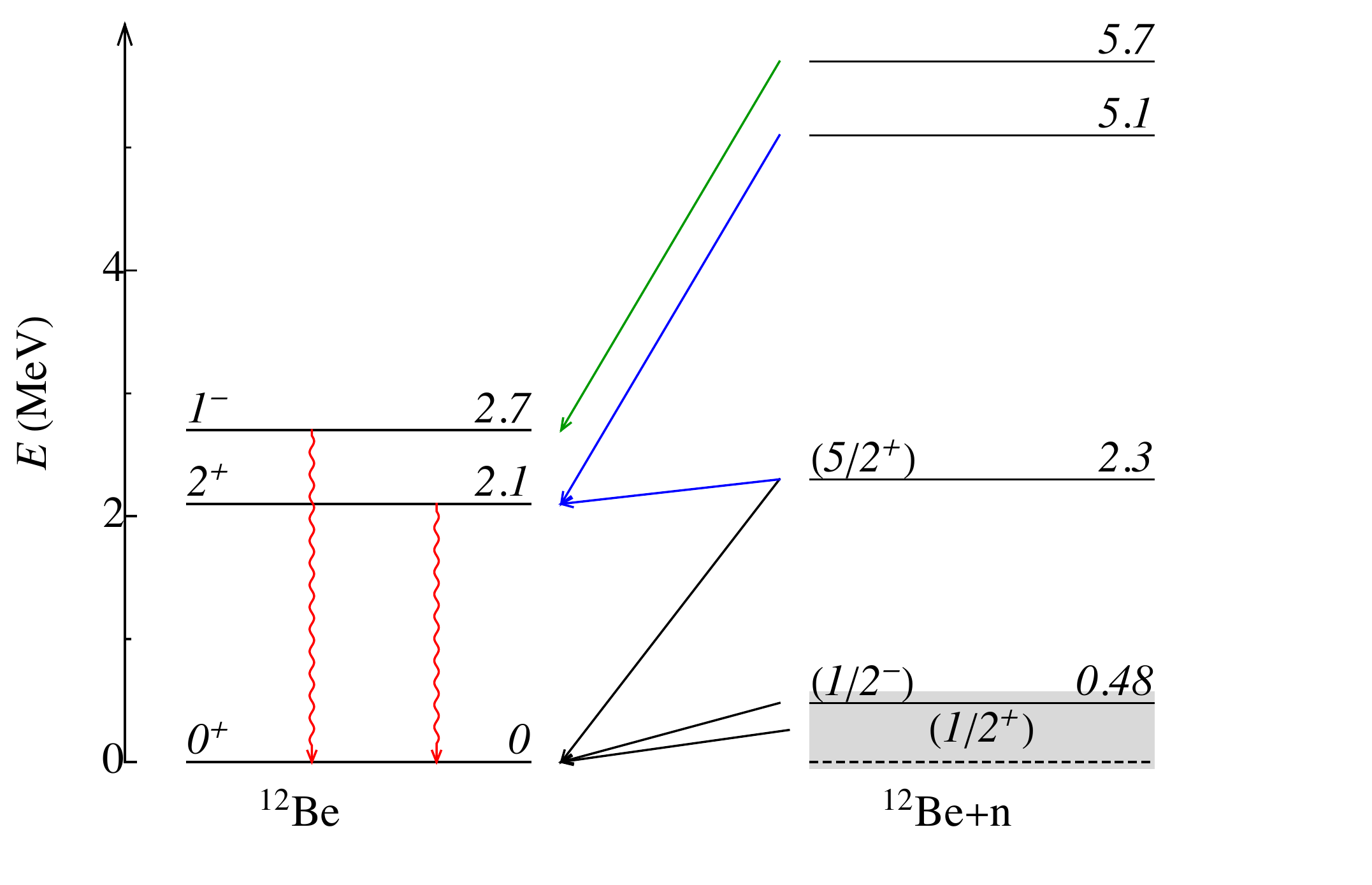}
     \caption{\label{f:ls} Partial level scheme based on the observed neutron-$^{12}$Be relative energy spectrum and gamma-neutron-$^{12}$Be coincidences. Transitions in the relative energy spectrum are represented by lines (black for transitions to the ground state of $^{12}$Be, blue and green for transitions populating $^{12}$Be(2$^+$) and $^{12}$Be(1$^-$), respectively). Gamma transitions are represented by the red wavy arrows. Energies are given in MeV.} 
 \end{center}
 \end{figure}

 Based on this, we built a partial level scheme presented in Fig.~\ref{f:ls}. Only the levels that can be clearly deduced from the present data are shown. The 2.3 MeV peak observed in the relative energy spectrum likely corresponds to the well-accepted 5/2$^+$ state in $^{13}$Be, whose tail may be responsible for the $\sim$0~MeV transition in coincidence with the  2$^+$ state in $^{12}$Be (as discussed, for instance, in Ref.~\cite{aks13b}). This, together with the spin-parity assignment of the lowest level at 0.48 MeV, will be further discussed in Sec.~\ref{sec:level4}, where we also analyze the information from the corresponding transverse momentum distributions.

\section{\label{sec:level4}Theoretical analysis}
\subsection{\label{ssec:level1}Three-body calculations}

In order to better understand the experimental results, we have performed structure calculations for  $^{14}$Be using a three-body model $(^{12}\text{Be}+n+n)$ within the hyperspherical formalism~\cite{Desc03,IJThompson04,MRoGa05}. Details on how the wave function is built can be found, for instance, in Ref.~\cite{JCasal16} and references therein. This consists in diagonalizing the Hamiltonian in an analytical transformed harmonic oscillator (THO) basis for three-body systems. The method has been previously applied with great success to describe direct reactions induced by three-body projectiles~\cite{JCasal15,casalplb17,gomezramosplb17,Arazi18}. 

Three-body calculations require, as input, the binary interactions between all constituents. For the $n$-$n$ potential, we employ the GPT tensor interaction~\cite{GPT}. This potential, although simpler than the more sophisticated AV18~\cite{av18}, CD-Bonn \cite{Bonn} or Reid93~\cite{reid93} interactions, reproduces $NN$ observables up to 300 MeV, so it is suitable for three-body structure calculations. In order to get a realistic description of $^{14}$Be, the $^{12}$Be-$n$ interactions needs to reproduce the properties of the unbound binary subsystem $^{13}$Be. From previous fragmentation and knockout experiments, it is mostly accepted that $^{13}$Be exhibits a low-lying $s$-wave state and a $d$-wave resonance around 2~MeV relative energy~\cite{tho00,lec04,sim07}. There are, however, large discrepancies in the interpretation of the $^{13}$Be spectrum from different experimental works~\cite{aks13b,rib18}, many of which are associated with the long-debated existence of a low-lying $p$-wave resonance and the contribution from excited $^{12}$Be components~\cite{kon10}. For this reason, we make use of different core-neutron potentials to study the sensitivity of the structure and reactions observables to the properties of $^{13}$Be.

In order to include some excited-core components in the description of $^{14}$Be, we parametrize the $^{12}$Be-$n$ interaction with a deformed Woods-Saxon potential with $l$-dependent central and spin-orbit terms. Following Ref.~\cite{tar04}, we introduce an effective quadrupole deformation parameter of $\beta_2=0.8$, and the $0^+$ ground state and the first $2^+$ excited state in $^{12}$Be are coupled by means of a simple rotational model~\cite{IJThompson04}. In this scheme, no other excited states of $^{12}$Be are included. Three-body calculations including also the 1$^-$ state in a consistent way are not available. As shown in Ref.~\cite{tar04}, despite the deformation the $s$-wave interaction still gives rise to a 1/2$^+$ virtual state in $^{13}$Be. The potential parameters $V_c^{(0,2)}$ and $V_{ls}$ are adjusted to fix the scattering length of this virtual state and to provide a 5/2$^+$ resonance just below the 2$^+$ excitation threshold in $^{12}$Be, i.e. at 2.11~MeV~\cite{Kelley17-12}. Note that, in this scheme, the 5/2$^+$ state may decay via $d_{5/2}$ neutrons to the ground state of $^{12}$Be, but also via $s_{1/2}$ to the 2$^+$ excited state, given its finite width. For simplicity, we start with a shallow $V_c^{(1)}$ potential, so no negative-parity resonances appear. This potential, labeled P1, produces a $1/2^+$ virtual state characterized by a large scattering length. Details are given in Table~\ref{tab:3b}. In addition, the calculations include a phenomenological three-body force to correct the position of the $0^+$ ground state to the experimental two-neutron separation energy of $^{14}$Be, i.e.~$S_{2n}=1.27$ MeV~\cite{Wang17}, which is kept fixed. Some properties of the resulting $^{14}$Be ground state are also given in Table~\ref{tab:3b}. We remind the reader that although the energy of a virtual state is strictly negative, it is customary to define a nominal positive energy as $E_s=\hbar^2/(2 \mu a^2)$ (with $a$ indicating the scattering length, see e.g.\ Chap.~2 of Ref.~\cite{Bla79}) to quantify the proximity of the virtual-state pole to the threshold. \\
Given the open debate about the presence of a low-lying $p$-wave resonance in $^{13}$Be, we consider another potential labeled P2. 
In this case, we increase the $p$-wave potential depth to produce a 1/2$^-$ resonance around the maximum of the $^{12}$Be-$n$ relative-energy distribution, while keeping a small scattering length for the $s$-wave state and the same $d$-wave resonance as with P1. With the adopted parameters, the scattering length of the 1/2$^+$ state is -9.2 fm, which corresponds to a nominal energy of 0.265 MeV. The computed energy and width of the 1/2$^-$ (5/2$^+$) resonance are 0.46 (1.96) and 0.40 (0.27) MeV. The resulting $^{14}$Be properties are also listed in Table~\ref{tab:3b}. 
It is worth noting that the P2 model produces a strong mixing between different-parity states, as shown in Table~\ref{tab:3b}. This gives rise to a marked dineutron character of the $^{14}$Be wave function, as opposed to potential P1 and in accord with the discussion in Ref.~\cite{Catara84}. 

In the next section, we will study the sensitivity of the $(p,pn)$ cross section to the structure properties of $^{13,14}$Be. For this purpose, we consider variations of potential P2 in which the $1/2^-$ and $5/2^+$ resonances are placed at different energies. These variations are labeled P3-5 and their details are also presented in Table~\ref{tab:3b}.

\begin{table*}[t]
\centering
\begin{tabular}{l|ccccc||ccc|c}
\hline
\hline
   & $a$  & $E(5/2^+)$  & $\Gamma(5/2^+)$  & $E(1/2^-)$  & $\Gamma(1/2^-)$  & $s$ & $p$ & $d$ & $2^+$ \\
\hline
P1 & $-40.1$  & 1.96 & 0.27 & - & - &  0.59 & 0.13 & 0.26 & 0.34\\
P2 & $-9.2$  & 1.96 & 0.27 & 0.46 & 0.40 &  0.19 & 0.62 & 0.18 & 0.22\\
P3 & $-9.2$  & 1.53 & 0.28 & 0.46 & 0.40 &  0.20 & 0.53 & 0.25 & 0.25\\
P4 & $-9.2$  & 2.10 & 0.19 & 0.46 & 0.40 &  0.17 & 0.68 & 0.13 & 0.19\\
P5 & $-9.2$  & 1.96 & 0.27 & 0.62 & 0.60 &  0.24 & 0.53 & 0.22 & 0.24\\
\hline
\end{tabular}
\caption{ \label{tab:3b} Scattering length $a$ (in fm) of the 1/2$^+$ virtual state in $^{13}$Be and energies and widths of the 5/2$^+$ and 1/2$^-$ resonances (in MeV) using the different core-neutron potentials P1-5. On the right, the resulting properties of the $^{14}$Be ground state: partial wave content for $L=0,1,2$ neutrons, 
and weight of the 2$^+$ core-excited components.
The two-neutron separation energy in $^{14}$Be is fixed to the experimental value of 1.27 MeV~\cite{Wang17}.
}
\end{table*}


\subsection{\label{ssec:level2}Reaction calculations}
We have compared the present $(p,pn)$ data with reaction calculations based on the so-called Transfer to the Continuum (TC) framework~\cite{AMoro15}, which was recently extended to describe processes induced by Borromean projectiles~\cite{gomezramosplb17}.  In this method, the differential cross section for the $(p,pn)$ reaction is obtained from the prior-form transition amplitude leading to the three-body continuum states for $p+n+^{13}$Be. The latter are expanded in a basis of $p$-$n$ states, which are conveniently discretized using a binning procedure akin to that employed in the continuum-discretized coupled-channels (CDCC) method \cite{Aus87}. Good convergence of the calculated observables (within 10\%) was attained using a set of energy
bins with a width $\Delta \epsilon_{pn}=15$ MeV, a maximum $p-n$ relative energy of $\epsilon_{pn}=210$ MeV and a maximum angular momentum and parity $j^\pi\leq 3^\pm$. 

The model considers a spectator/participant scenario, in which the incident proton  is assumed to remove one of the valence neutrons without modifying the state of the remaining  $^{13}$Be ($^{12}{\rm Be} + n$) subsystem. This is consistent with QFS conditions. Under this assumption, the $^{14}$Be structure enters through the overlap functions between the initial state (the $^{14}$Be g.s.) and the final $^{13}$Be states so that the cross section for  different configurations of $^{13}$Be (defined by their energy $E_{n-^{12}{\rm Be}}$ and angular momentum and parity $J_T^\pi$)
can be computed independently. 

Important ingredients are also the proton-neutron and nucleon-$^{13,14}$Be interactions. For the former, following previous applications of the method \cite{AMoro15,gomezramosplb17}, we adopt  the Reid93 parametrization \cite{reid93}, which provides an accurate description of the proton-neutron cross sections and phase-shifts up to 350~MeV. For proton-$^{14}$Be and proton/neutron-$^{13}$Be interactions, which take into account the distortion and absorption of the incoming proton and of the outgoing nucleons, we employ the optical potentials computed from the Dirac phenomenological parametrization \cite{Ham90,Coo93}.

\begin{figure}[hbt!]
\centering
\includegraphics[width=0.8\linewidth]{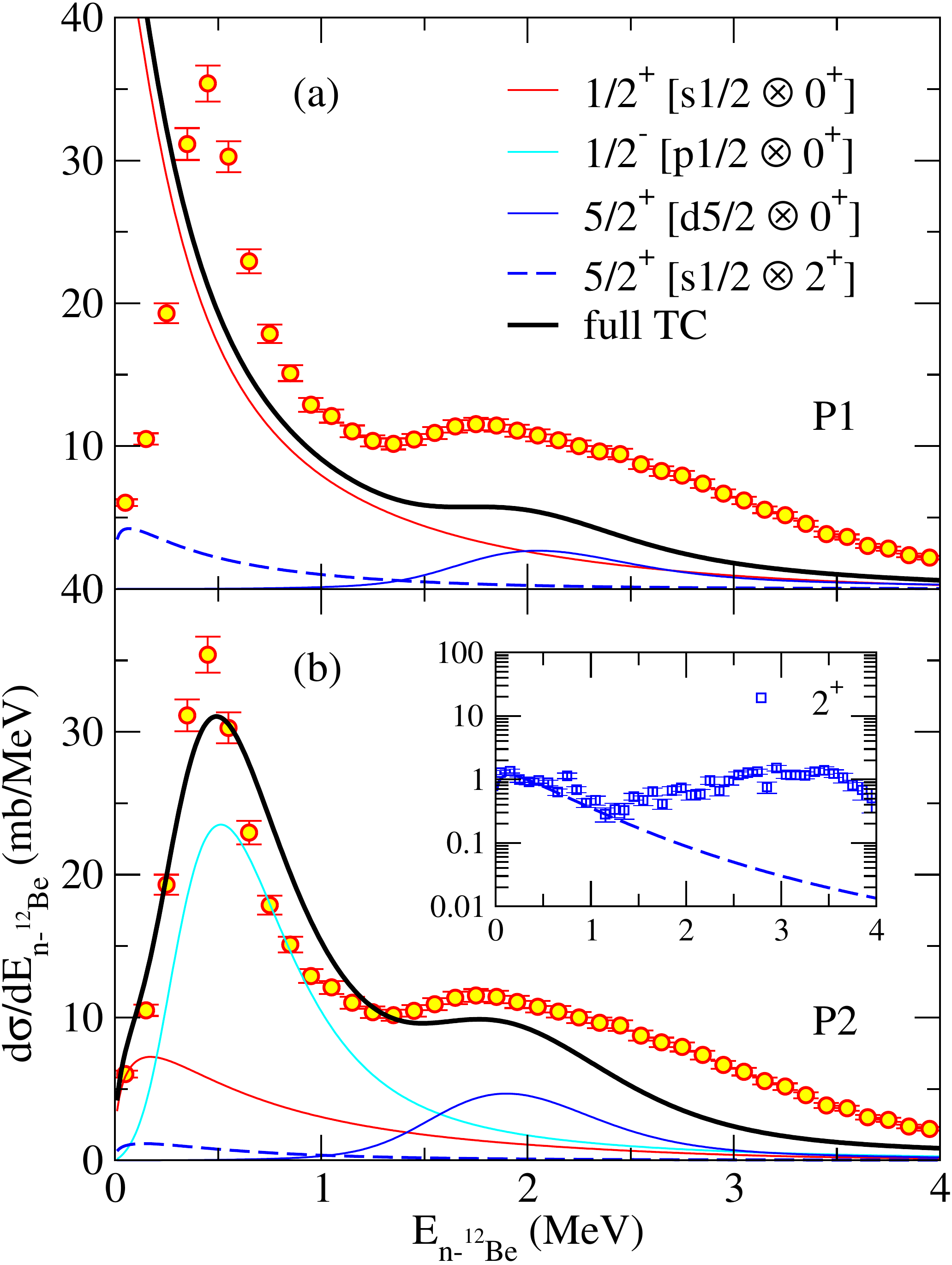}
\caption{$^{13}$Be relative-energy spectrum using the (a) P1 and (b) P2 potentials. Results are shown after convoluting the theoretical lineshapes with the experimental resolution function. The inset shows the relative energy spectrum measured in coincidence with the $^{12}$Be(2$^+$) decay transition, compared to the calculated core-excited component. See text for details.}
\label{fig:13Be-P1_P2}
\end{figure}

The $^{13}\text{Be}={^{12}}\text{Be}+n$ relative-energy spectrum obtained by using the P1 core-neutron potential and convoluting with the experimental resolution from the simulation (determined as explained in Sec.~\ref{sec:level2}) is shown in Fig.~\ref{fig:13Be-P1_P2}a. Note that these reaction calculations provide absolute cross sections, so no fitting or scaling is carried out. The different contributions are labeled $J_T [L_J\otimes I]$, where $J_T$ is the total $^{13}$Be angular momentum resulting from coupling the single-particle configuration $L_J$ with the spin $I$ of $^{12}$Be. Trivially, since the spin of $^{14}$Be is $0^+$, $J_T$ equals $j_2=[l_2 \otimes s_2]$, the angular momentum of the removed nucleon. In this figure, only the leading components are shown, together with the total cross section. In this way, the contributions from $I=0$ and $I=2$ can be separated. The cross section using P1 overestimates the experimental data at low relative energies due to the large scattering length of the $s$-wave virtual state. Moreover, the maximum of the cross section around 0.5~MeV is not reproduced. We have checked that variations of the position of the 1/2$^+$ virtual state do not improve the agreement. 

The results using the P2 potential are shown in Fig.~\ref{fig:13Be-P1_P2}b. Again, for clarity only the leading terms are presented. Notice that, although the width of the 5/2$^+$ state in the present model is smaller than that of the 1/2$^-$ (see Table~\ref{tab:3b}), its contribution to the relative-energy spectrum becomes broader due to the energy resolution. Note also that the contribution of the $2^+$ state is small in spite of the significant core-excited component in these models. This is because the nominal energy of the 5/2$^+$ state, which collects most of the core-excitation weight in the $^{14}$Be wave function, lays below the $^{12}$Be($2^+$) excitation threshold. 
The small $5/2^+[s_{1/2}\otimes 2^+]$ contribution is consistent with the analysis from gamma coincidences. The comparison between this component and the experimental data in coincidence with the $^{12}$Be(2$^+$) decay transition is presented in the inset of Fig.~\ref{fig:13Be-P1_P2}b, where we can see that the calculations describe the data at low energies quite well. The disagreement at energies above $\sim 1.5$ MeV might indicate that the present three-body calculations are missing some high-lying state which can also decay via $^{12}$Be($2^+$), as suggested in Fig.~\ref{f:ls}. As for the full spectrum, Fig.~\ref{fig:13Be-P1_P2}b shows that the low-energy peak can be described reasonably well by the 1/2$^-$ resonance using the P2 potential. The theoretical distribution is somewhat broader than the experimental data, so the calculation overestimates the measurements between $\sim$0.6-1 MeV. This might be an indication that the $p$-wave content obtained with the P2 potential is perhaps too large. In addition to the underestimation at large relative energies, this suggests that there might be missing components in the wave-function expansion, in particular those coming from the decay of other states in $^{13}$Be via $^{12}$Be($2^+$), or the coupling to other excited states of $^{12}$Be. Calculations including these features are not available yet.

\begin{figure}[hbt!]
\centering
\includegraphics[width=0.8\linewidth]{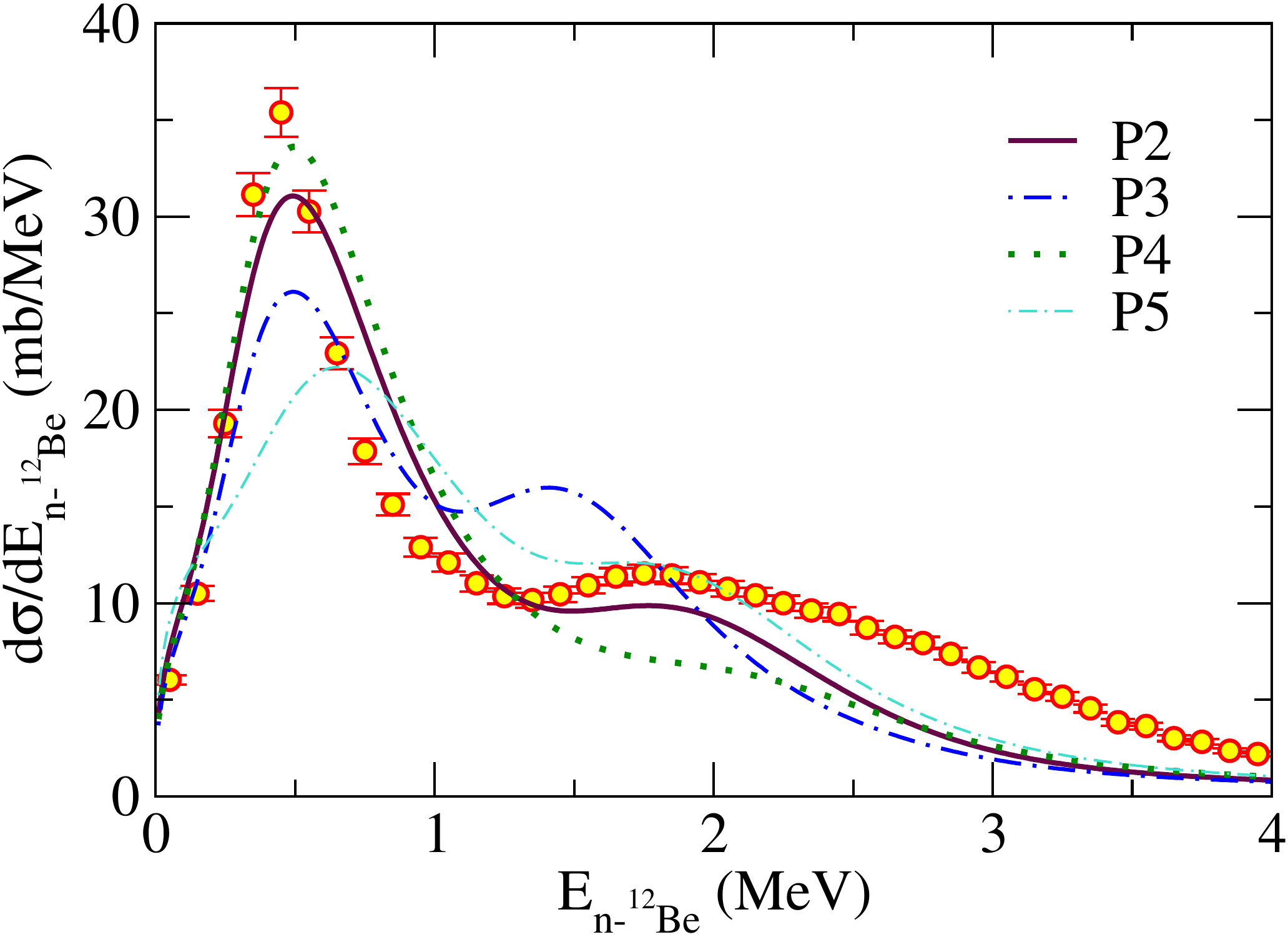}
\caption{Results for the $^{13}$Be relative-energy spectrum with potentials P2-5. Only the total calculations are presented.}
\label{fig:13Be-PX}
\end{figure}

To test the sensitivity of the relative-energy spectrum to the specific features included in the potential, in Fig.~\ref{fig:13Be-PX} we compare the P2 calculations with three additional models. As shown in Table~\ref{tab:3b}, P3 and P4 are obtained by placing the 5/2$^+$ resonance of $^{13}$Be at lower or higher energies, respectively, while P5 involves a variation on the position of the 1/2$^-$ state. Note that a difference in the position of the relevant levels changes also the weight of the different components in the ground state of $^{14}$Be. It is shown that the best agreement up to 2 MeV relative energy is achieved using the potential P2.


\begin{figure}[hbt!]
\centering
\includegraphics[width=1\linewidth]{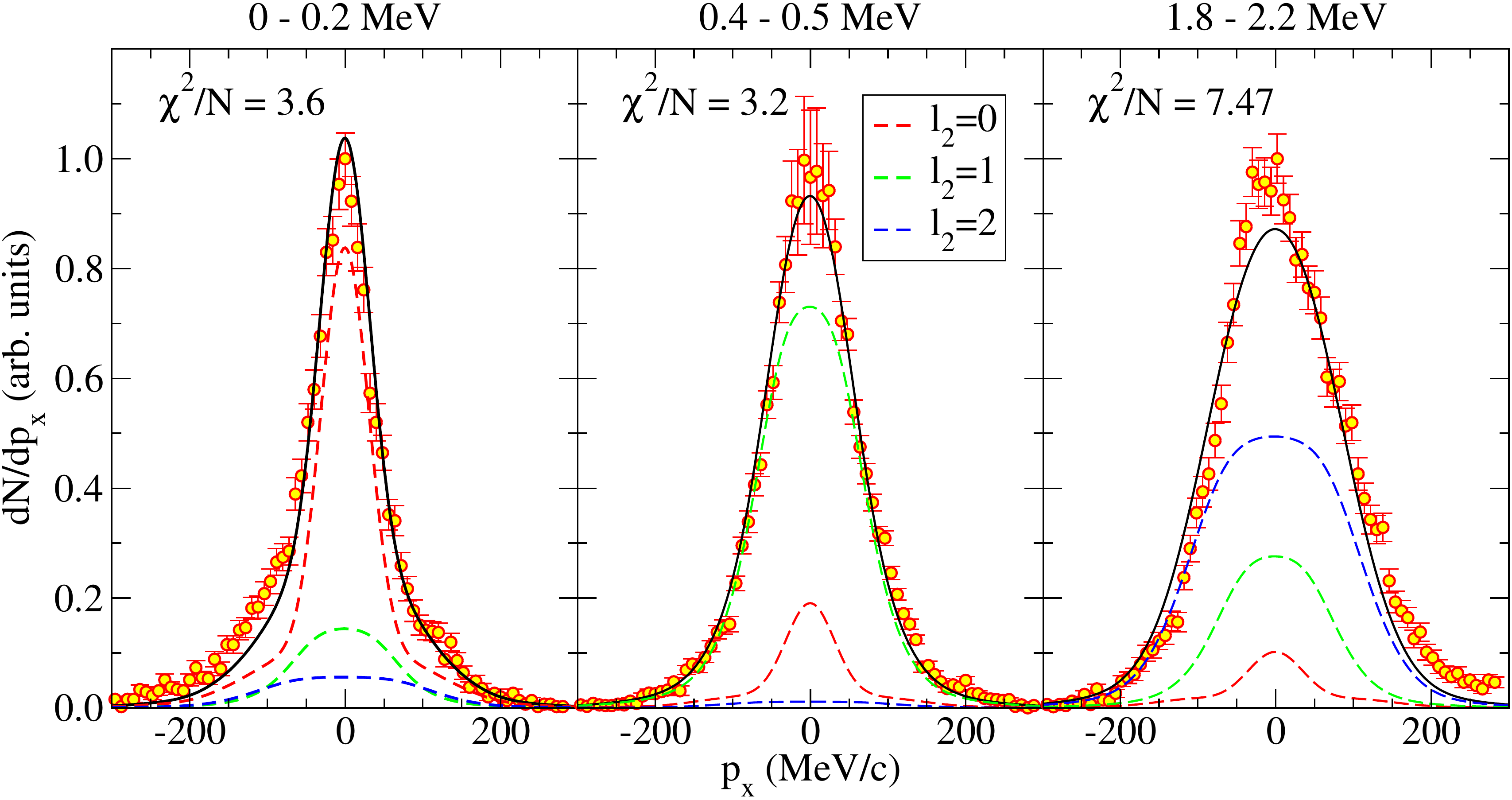}
\caption{Experimental transverse momentum distributions at 0-0.2, 0.4-0.5 and 1.8-2.2 MeV relative energy with P2 potential. The solid black line is the total TC result, convoluted with the experimental resolution and globally rescaled through a $\chi^2$ fit. Dashed lines are the contributions corresponding to removal of a neutron from a $s$- (red), $p$- (blue) or $d$-wave (green). }
\label{fig:momdist}
\end{figure}

The $^{13}$Be structure can be further studied from the transverse momentum distributions of the knocked-out neutron. The comparison between the present calculations, using potential P2, and the experimental momentum distributions is presented in Fig.~\ref{fig:momdist}, for three different relative-energy bins: 0-0.2, 0.4-0.5 and 1.8-2.2 MeV. Calculations have been convoluted with the experimental resolution of $\sim$~39 MeV/c (FWHM) obtained from the direct beam measurement. The contribution from momentum resolution of the neutron was checked via the simulation and found to be negligible. The overall normalization of the total theoretical distribution with respect to the data has been adjusted to obtain the best $\chi^2$ fit. 
Individual contributions from the different orbital angular momenta of the knocked-out neutron are also presented. The relative weights of these contributions are fixed by the structure and reaction calculations and not via $\chi^2$ fit. Note that, due to the non-zero spin of the $^{12}$Be core in its excited components, the orbital angular momentum $l_2$ of the removed neutron is not necessarily the same as the one of the valence neutron in $^{13}$Be.
Overall, it is found that the width of the momentum distributions is well reproduced. 
In particular, we can describe the data at 0.4-0.5 MeV with a dominant $p$-wave contribution. In Ref.~\cite{aks13b}, an $s$-wave resonance (or a combination of two overlapping $s$-wave resonances) was proposed to explain the peak in the $^{13}$Be spectrum. 

\begin{figure}
\centering
\includegraphics[width=0.75\columnwidth]{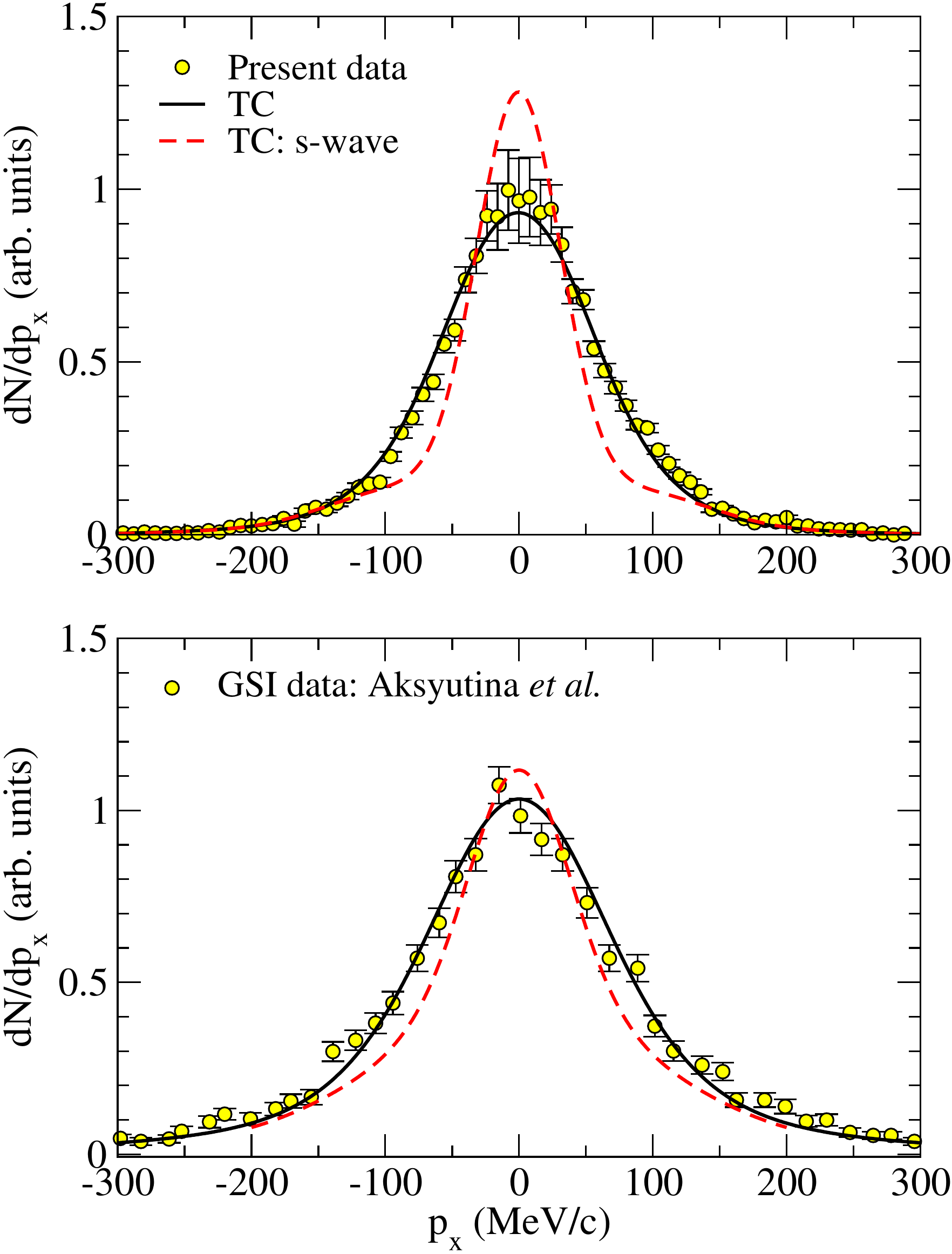}
\caption{Transverse momentum distribution at 0.4-0.5~MeV relative energy. The top and bottom panels are for the present experiment and from the GSI data at 304 MeV/nucleon with resolution $\sim$72 MeV/c (FWHM) \cite {aks13b}, respectively. The black solid line is the total P2 result, while the  red dashed line corresponds to a $\chi^2$ fit assuming a pure $s$-wave distribution.}
\label{fig:momdist2}
\end{figure}

To test the assumption of the $s$-wave resonance suggested in Ref.~\cite{aks13b}, in the top panel of Fig.~\ref{fig:momdist2} we have performed a $\chi^2$ fit of the momentum distribution for the 0.4-0.5 energy bin retaining only the $s$-wave contribution in our calculations with the P2 potential. In this case, the resulting width is clearly too small. A similar conclusion  is achieved if the analysis is performed over the data from Ref.~\cite{aks13b}, which is shown in the bottom panel. In this latter case, our full calculation with a dominant $p$-wave component reproduces very well the experimental momentum distribution, whereas the pure $s$-wave assumption gives again a too narrow distribution. Our analysis shows that the peak observed in the $^{13}$Be relative-energy spectrum at $E_r \sim 0.5$~MeV, populated in the $^{14}$Be($p,pn)$ reaction, is most likely dominated by the $p$-wave contribution.
This assignment is in disagreement with that of Ref. \cite{aks13b}. We believe this discrepancy can be understood as due to the inherent uncertainties in the $\chi^2$ procedure used in Ref. \cite{aks13b} to assign the orbital momentum, as well as the differences in structure and reaction model.

\section{\label{sec:level5} Conclusions}
We have presented a high statistics measurement of the spectroscopy of $^{13}$Be via invariant mass, including the measurement of $^{12}$Be core excited states which decay via gamma rays. We clearly observed for the first time the contribution of both $^{12}$Be(2$^+$, 1$^-$) states in $^{13}$Be populated via $(p,pn)$ reaction. Their contribution to the $^{13}$Be relative-energy spectrum is small. A still missing information is the contribution of the isomeric $^{12}$Be(0$^+_2$) core-excited state, that will demand a dedicated measurement. 
A key and novel aspect of our analysis consists in calculating, for the first time, the relative energy cross section and momentum distribution using a well-founded reaction framework and a realistic three-body model of $^{14}$Be that incorporates $^{12}$Be(2$^{+}$) excitations, thus avoiding the common procedure of extracting individual angular momentum  components from a fit, a technique that becomes more ambiguous in the case of complex spectra with overlapping structures as the one of $^{13}$Be.
This analysis permitted to pin down the dominant $\ell=1$ contribution of the resonant peak observed in the low-lying spectrum, in agreement with \cite{kon10,rib18} and at variance with the conclusions of Refs.~\cite{aks13a, aks13b}, which assigned a dominant $\ell=0$ to this peak.

Additional observables, such as the distribution of the opening angle between the momenta of $^{13}$Be and the removed neutron in the final state, may help to shed light on the structure of $^{14}$Be and hence $^{13}$Be. The interpretation of such observables, as extracted from the present experiment, will be the subject of an upcoming publication.  An improvement of the three-body model, e.g. the inclusion of other $^{12}$Be excited state, proper antisymmetrization of valence neutrons and a better treatment of Pauli principle, will increase the capability of theory to capture the features of the experimental spectrum.



\section*{Acknowledgements}
This work has been supported by the European Research Council through the ERC Starting Grant No. MINOS-258567. A.C. acknowledges Y. Kikuchi and K. Ogata for fruitful discussions, and N.Paul for careful proofreading. M.G.R., A.M.M.\  and J.C.\ are partially supported  by the Spanish Ministerio de Ciencia, Innovaci\'on y Universidades and FEDER funds (projects FIS2014-53448-C2-1-P and FIS2017-88410-P)  and by the European Union's Horizon 2020 research and innovation program under grant agreement No.\ 654002. J.G., F.M.M. and N.A.O. acknowledge partial support from the Franco-Japanese LIA-International Associated Laboratory for Nuclear Structure Problems as well as the French ANR14-CE33-0022-02 EXPAND.

\section*{References}
\bibliography{biblio2}

\end{document}